\documentclass[aps,reprint, twocolumn]{revtex4}
\usepackage{graphicx}
\usepackage{amsmath}
\usepackage{amssymb}
\usepackage{mathrsfs}
\usepackage{amsthm}
\usepackage{hyperref}
\usepackage{tikz}
\usepackage{pgfplots}
\usetikzlibrary{intersections, pgfplots.fillbetween, shapes, arrows,decorations.pathmorphing}
\tikzstyle{snakeline} = [decorate, decoration={pre length=0.1cm,
                         post length=0.1cm, snake, amplitude=.5mm,
                         segment length=2mm},thick, red, ->]

\newcommand{\bg}{\bar g}
\newcommand{\pa}{\partial}

\newcommand{\Tr}{{\rm Tr}}

\renewcommand{\a}{\alpha}
\renewcommand{\b}{\beta}
\renewcommand{\c}{\gamma}

\renewcommand{\d}{\delta}

\newcommand{\s}{\sigma}

\renewcommand{\t}{\theta}

\newcommand{\la}{\lambda}

\newcommand{\ud}{\mathrm{d}}          
\newcommand{\ue}{\mathrm{e}}    
\newcommand{\be}{\begin{equation}}
\newcommand{\ee}{\end{equation}}
\newcommand{\bea}{\begin{eqnarray}}
\newcommand{\eea}{\end{eqnarray}}
\newcommand{\ba}{\begin{align}}
\newcommand{\ea}{\end{align}}

\newcommand{\nn}{\nonumber\\}

\newcommand{\cN}{\mathcal{N}}

\newcommand{\cO}{\mathcal{O}}

\newcommand{\cD}{\mathcal{D}}

\newcommand{\cZ}{\mathcal{Z}}

\begin{document}

\title{  Gravitational realization of magnons in a ferromagnetic spin chain}

\author{ Juan  Diego Chang}
\email{jdiegochang@ecfm.usac.edu.gt}
\author{ Rodrigo de Le\'on Ard\'on}
\email{rdeleon@ecfm.usac.edu.gt}
\author{ Juan  Ponciano}
\email{japonciano@ecfm.usac.edu.gt}
\author{ Giovanni Ram\'irez}
\email{ramirez@ecfm.usac.edu.gt}

\affiliation{
 Instituto de Investigaci\'on en Ciencias F\'isicas y Matem\'aticas,
Escuela de Ciencias F\'isicas y Matem\'aticas (ECFM),
Universidad de San Carlos de Guatemala, Guatemala.}

\date{ \today}

\begin{abstract}
A gravitational model of  magnons in thermal equilibrium with a ferromagnetic spin chain is developed in a phenomenological bottom-up approach. A large Schwarzschild-AdS  black hole background is used as the thermal reservoir and the magnon dynamics is obtained by  scalar fields and branes in the bulk.  The key feature of this model is that the coupling of the spin chain is related with the radial position in which the brane is located. We further study a ferromagnetic spin chain with a competing interaction and find that the couplings are related by the difference of positions of the branes. 
We show how to obtain the model from a weak limit of a  dynamical gravitational system. This allows us to embed the model into a holographic system. The couplings can be related to entanglement entropy at finite temperature of the CFT since the turning point of minimal  surfaces coincides with the position of the branes. 
The difference of  entropy is used to define a notion of \emph{distance} between the chain couplings.
\end{abstract}

\maketitle

\section{Introduction}

In ferromagnetism the temperature of the thermal reservoir must be below the Curie temperature in order to obtain spontaneous magnetization.  Moreover, at zero temperature the ferromagnetic spin chain is in its ground state and the spontaneous magnetization reaches its saturation value. For small temperatures,  spin-wave excitations occur and the saturation magnetization receives finite temperature corrections \cite{kittelbook,AAFerro,AshMermin,sachdev_2011}.  Spin-wave theory tells us that these corrections can be systematically studied in the  large spin limit (semiclassical limit) by an effective theory of magnons (the quanta of spin waves). 

Ferromagnetism corresponds to a system  in which  rotational symmetry is spontaneously broken. Spontaneous magnetization arises from a vacuum expectation value  and the Goldstone modes correspond to magnons. At finite temperature, spontaneous symmetry breaking of continuus symmetries cannot be realized in one or two spatial dimensions \cite{PhysRevLett.17.1133,1973CMaPh..31..259C}. Therefore, the effective theory of magnons must be defined in $p=3$ spatial dimensions. For low momenta, the leading order contribution of magnons to the excitated states of the spin chain Hamiltonian can be represented by a $O(3)$ NL$\s$ model of scalar fields  in the continuum limit \cite{Affleck:1988mua,Rao_1997,sachdev_2011}.  

Taking in consideration the phenomenology of ferromagnetism and spin-wave theory. The proposed gravitational model considers a \emph{large} Schwarzschild-AdS  black hole (SAdS-BH) as the thermal reservoir with its temperature below the Curie point. In fact this black hole has been exploited in the AdS/CFT correspondence \cite{Maldacena:1997re,Gubser:1998bc,Witten:1998qj} by relating the gravity side to a high temperature phase of a dual field theory. The correspondence is holographic; the  spacetime background of the field theory is identified with the boundary of AdS. Motivated by condensed matter applications of the AdS/CFT correspondence, see \cite{Hartnoll:2016apf, Nastase:2017cxp} for a review, we are interested in the dynamics of scalar fields and fixed $p$-branes on a \emph{large} SAdS-BH background (in the planar limit) in order to study  magnons of a quantum ferromagnetic spin chain in thermal equilibrium. 

The magnons correspond  to scalar fields/brane system in the black hole background. In a holographic context, ferromagnets have been already studied in the literature \cite{Iqbal:2010eh,Cai:2014oca,Yokoi:2015qba,Cai:2015jta}. A more abstract relation with ferromagnetism was realized in \cite{Minahan:2002ve} (see also \cite{Beisert:2003tq, Kruczenski:2003gt, Kruczenski:2004kw, Zarembo:2004hp,  Nastase:2017cxp}). The computation of 1-loop anomalous dimension for certain $\cN = 4$ super Yang Mills operators  can be map to the diagonalization of a spin $\frac{1}{2}$ ferromagnetic Heisenberg chain. Magnons correspond to impurities (other fields) that propagate  in the chain. The dispersion relation of these magnons were computed from a semiclassical rotating string in $AdS_5\times S^5$ in \cite{Hofman:2006xt}.

 In a phenomenological approach, Yokoi et al. \cite{Yokoi:2015qba} developed the dictionary between a ferromagnetic system with a holographic system.  The field theory content of their model follows from an action invariant under a gauged $SU(2)$ and global $U(1)$. The ordered state is achieved since the $SU(2)$ is spontaneously broken to $U(1)$ via the ``Mexican hat'' scalar potential. 

Here we also follow a phenomenological approach but realize ferromagnetism in a different way. We are interested in the scalar fields in the adjoint of the $SU(2)$. No external current is considered and we focus only on the magnon contribution on the saturation magnetization. The magnon NL$\s$ model is obtained from the on-shell action of the scalar fields evaluated at large limit of the radial holographic coordinate in the presence of a $3$-brane with negative tension. The NL$\s$ model coupling is found to be strong and non linearity in the $\s$-model is achieved via the brane. Its effect is to compactify the flat target space (after adding a \emph{north pole}) of the scalar fields to a sphere. The key feature of this model is that the coupling of the spin chain, $J$, which determines the nature of the ground state of the chain is related with the radial position in which the 3-brane is located.

This setup provides a framework to study ferromagnetic spin chains with competing interactions. 
These type of systems can be obtained by a suitable addition of independent scalar fields with their respective $3$-branes. The couplings are found to be related to the positions of the branes and their absolute value behaves as local fiducial temperatures, i.e.  temperatures measured by an observer at a fixed distance from the black hole. It is found that the presence of the second brane allows the possibility of other state that is a local maximum of the energy instead of a solely ferromagnetic ordered state.

\begin{figure}[ht!]
\centering
\begin{tikzpicture}
\node at (2,4) [rectangle,draw] (qsc) {Quantum spin chain}; 
\node at (2,2) [rectangle,draw] (lpw){Linear spin-waves}; 
\node at (4,0) [rectangle,draw] (gads){SAdS-BH}; 
\node at (0,0) [rectangle,draw] (cft){CFT};
\node at (2,3) [rectangle,draw] (scl){Semiclassical limit};
\draw[thick] (qsc)--(scl);
\draw[->, thick]  (scl)--(lpw);
\draw[<->, thick] (lpw)--(gads);
\draw[<->, thick] (gads)--(cft);
\draw[->, thick] (cft)--(lpw);
\end{tikzpicture}
\caption{Relation between theories at finite temperature. }
\label{theoriesrel}
\end{figure}
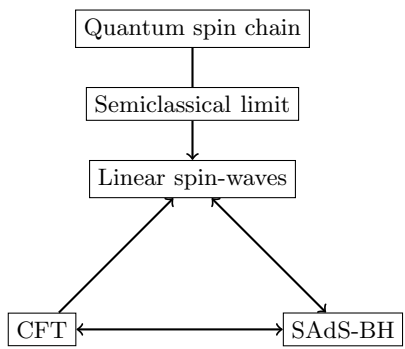

We also show how to obtain the model from a weak limit of a dynamical gravitational system. This allows us to embed the model into a holographic system. The couplings can be related to entanglement entropy at finite temperature since the turning point of minimal  surfaces coincides with the position of the branes.
The overall relation of the theories is depicted in Figure \ref{theoriesrel}. The figure shows how the embedding into a holographic theory is realized. 
The consequences of these relations are summarized in a \emph{thermodynamic diagram} depicted in Figure \ref{thermo}.  In that diagram, an \emph{isotermal} curve is at constant Hawking temperature in the CFT and an \emph{adiabatic} curve is realized in the bulk.  Finally, the difference of entropy is used to define a notion of \emph{distance} between the chain couplings.

In order to provide a clear exposition to the possible communities interested in this work,  the paper briefly reviews fundamental concepts in gravitational physics and spin chain physics. It is organized as follows. In section II we summarize relevant properties of the large SAdS-BH in the planar limit and the dynamics of a scalar field in such background. Section III is devoted to reviewing spin-wave theory of the quantum ferromagnetic XXX Heisenberg spin chain, its connection with the gravitational dual and constructing  the gravitational dual model of magnons of a ferromagnetic spin chain with a competing interaction. Finally, section IV contains the summary and discussion.

\section{Large SAdS  black hole}
The geometry  and thermodynamics of a SAdS-BH \cite{Hawking:1982dh,Brynjolfsson:2008uc,Hemming:2007yq,Hubeny:2009rc,Kubiznak:2016qmn} compared with an asymptotically flat Schwarzschild black hole (S-BH) differ in several interesting ways. The evident geometrical difference corresponds to the behaviour in the infrared regions. Contrary to the S-BH, the SAdS-BH is regulated in the IR limit since the global AdS geometry acts as a confining box. This has important thermodynamic consequences. For example, the local fiducial temperature measured by  fiducial observers in a S-BH approaches the Hawking temperature far away from the black hole and diverges in the horizon. For the SAdS-BH, local fiducial temperature vanishes at the boundary. 

Another relevant geometrical difference between the black holes is that SAdS-BH has two characteristic length scales: the radius of AdS, $L$, and the horizon radius $r_+$. Therefore, the Hawking temperature  scales differently for particular values of $L$ and $r_+$. The SAdS-BH splits into two branches: \emph{large} black hole ($r_+\gg\sqrt{\frac{p-1}{p+1}}L$) and \emph{small} black hole ($r_+\ll\sqrt{\frac{p-1}{p+1}}L$), where $p$ denotes the spatial dimensions. The former is known to follow the Stefan-Boltzmann law; it has positive heat capacity and is thermodynamically stable. This should be contrasted with the asymptotically flat S-BH; it has negative heat capacity  and  does not reach a stable equilibrium. 

The line segment of a Euclidean $AdS_{p+2}$ static spherically symmetric black hole is given by
\be
\ud s^2=f(r)\ud\tau^2+\frac{\ud r^2}{f(r)}+r^2\ud\Omega^2_p,  
\label{ESAdS1}
\ee
with 
\be 
f(r)=1+\frac{r^2}{L^2}-\left(\frac{r_+}{r}\right)^{p-1}\left(1+\frac{r_+^2}{L^2}\right).
\ee
The boundary is located at $r\to\infty$ and for $r_+=0$ the metric reduces to the static patch of hyperbolic space. If we consider the scaling $\tau \to \la\tau$, $r\to r/\la$ and $r_+\to r_+/\la$, the metric is not invariant, which reflects the inequivalence of AdS black holes with spherical horizon at different radii. The Hawking temperature is
\be
T_H=\frac{1}{4\pi}\left[(p+1)\frac{r_+}{L^2}+(p-1)\frac{1}{r_+}\right].
\ee
It has a local minimum at $r_+^*=\sqrt{\frac{p-1}{p+1}}L$ and therefore $T_H\geq \frac{1}{2\pi L}\sqrt{(p-1)(p+1)}$. The SAdS-BH can be classified into two branches: \emph{large} $r_+\gg r_+^*$ and \emph{small}  $r_+\ll r_+^*$. One can see that the temperature of the \emph{large} SAdS-BH grows linearly with $r_+$ and the \emph{small} SAdS-BH decays as $1/r_+$. The former case is relevant to the the AdS/CFT correspondence and the latter corresponds to the usual temperature dependence of a Schwarzschild black hole horizon radius. If we consider the large limit $r_+\gg r_+^*$ and a rescaling of the coordinates, the metric is invariant if and only if the horizon is planar, i.e. $r^2\ud\Omega^2_p\approx (r/L)^2\ud\mathbf{x}_p^2$. These black holes are referred to planar AdS black hole \footnote{Black holes with planar horizon $\mathbb{R}^p$ are called black $p$-branes in String theory. For example, in the AdS/CFT correspondence at finite temperature one considers a black D3-brane solution and the near horizon limit. As a result one obtains a planar  $AdS_5$ black hole times $S^5$, see \cite{Natsuume:2014sfa} for a review.} and its metric is given by
\be
 \ud s^2=\frac{r^2}{L^2}\left(h(r)\ud\tau^2+\ud\mathbf{x}^2_p\right)+\frac{L^2}{r^2}\frac{\ud r^2}{h(r)}, 
 \label{ESAdS2}
\ee
with
\be
 h(r)=1-\left(\frac{r_+}{r}\right)^{p+1}.
\ee
We can conclude that planar horizon AdS black holes with different horizon radii are equivalent by scale transformations. This equivalence, in physical terms, implies that there is no local measurement of the temperature and its functional form can be determined simply by scaling and dimensional analysis which  can check this from
\be 
T_H=\frac{(p+1)}{4\pi}\frac{r_+}{L^2}.
\ee
From now on, we will use the planar horizon approximation of the SAdS-BH. The energy and entropy density of the planar black hole are 
\be
\varepsilon = \frac{p}{16\pi G_{p+2} L} \left(\frac{r_+}{L}\right)^{p+1}, \quad  \varsigma = \frac{1}{4 G_{p+2}} \left(\frac{r_+}{L}\right)^{p},
\label{densities}
\ee
where $G_{p+2}$ is the $(p+2)$-dimensional Newton's constant. One particular thermodynamic property of the planar SAdS-BH is that it follows the Stefan-Boltzmann law. Together with its stability, this justifies the consideration of this black hole as a thermal reservoir. 
\subsection{Scalar field in the large SAdS-BH background}
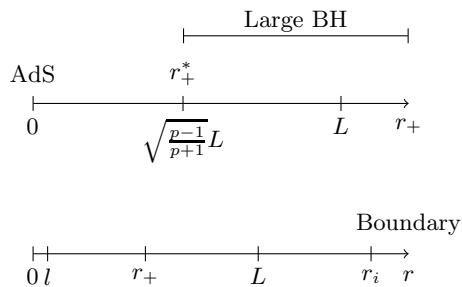
\begin{figure}[ht!]
\centering
\begin{tikzpicture}
\node at (0,-0.3) {0};
\node at (5,-0.3) {$r$};
\node at (0,1.7) {0};
\node at (5,1.7) {$r_+$};
\node at (0,2.4) {AdS};
\node at (2,2.4) {$r_+^*$};
\node at (2,1.5) {$\sqrt{\frac{p-1}{p+1}}L$};
\node at (4.1,1.7) {$L$};
\node at (0.2,-0.3) {$l$};
\node at (1.5,-0.3) {$r_+$};
\node at (3,-0.3) {$L$};
\node at (4.5,-0.3) {$r_i$};
\node at (5,0.4) {Boundary};
\draw (2,1.9) -- (2,2.1);
\draw (4.1,1.9) -- (4.1,2.1);
\draw (0.2,-0.1) -- (0.2,0.1);
\draw (1.5,-0.1) -- (1.5,0.1);
\draw (3,-0.1) -- (3,0.1);
\draw (4.5,-0.1) -- (4.5,0.1);
\node at (3.5,3.1) {Large BH };
\draw[|-|] (2,2.9) -- (5,2.9);
\draw[|->] (0,0) -- (5,0);
\draw[|->] (0,2) -- (5,2);
\end{tikzpicture}
\caption{The characteristic scales of the system are depicted. The top line shows the regions of the horizon radius $r_+$ for \emph{small} and \emph{large} black hole, as well as the point of minimum temperature $r_+^*$. The bottom line shows the characteristic lengths of the system.}
\label{relscales}
\end{figure}

The wave equation of a dimensionless and massless scalar field $\Phi$ in the background given by Eq. \eqref{ESAdS2} is
\begin{multline}
 h\Phi'' +\left(h'+(p+2)\frac{h}{r}\right)\Phi'\\+\frac{L^4}{r^4}\left(\frac{1}{h}\ddot{\Phi}+\pa_i\pa^i\Phi\right)=0,
 \label{wave.eq}
\end{multline}
where $ \Phi' $ stands the partial derivative respect to $r$ and $\dot\Phi$ with respect to $\tau$. Since the horizon is planar and at finite temperature we have $\tau\sim\tau+\b$ and the scalar field can be expanded in terms of Matsubara and Fourier modes:
\be
\Phi(r,\tau,\mathbf{x})=l^p\sum_{n=-\infty}^{\infty}\int\frac{\ud^p\mathbf{k}}{(2\pi)^p}\ue^{i\frac{2\pi n \tau}{\b}}\ue^{i\mathbf{k}\cdot\mathbf{x}}\tilde{\Phi}_{n,\mathbf{k}}(r), 
\label{expansionphi}
\ee
where $l$ is a characteristic  length scale of the system that provides a UV cutoff. Therefore, the geometry suggests that the field can be decomposed as
\be
 \Phi(r,\tau,\mathbf{x})=\Phi_0(r,\mathbf{x})+\Theta(r,\tau,\mathbf{x}),
\label{expansionphi2} 
\ee
where $\Phi_0$ is static and corresponds to the case $n=0$. The field $\Theta$ involves the rest of the modes. The wave equation is obtained  after substituting the expansion \eqref{expansionphi} into Eq. \eqref{wave.eq}. Thus, it is reduced to
\begin{multline}
 h\tilde\Phi_{n,\mathbf{k}}'' +\left(h'+(p+2)\frac{h}{r}\right)\tilde{\Phi}'_{n,\mathbf{k}}\\
 -\frac{L^4}{r^4}\left(\frac{1}{h}\left(\frac{2\pi n}{\b}\right)^2+\mathbf{k}^2\right)\tilde{\Phi}_{n,\mathbf{k}}=0.
 \label{radial.wave.eq}
\end{multline}
This radial equation is subject to boundary conditions. Near the AdS boundary the solution is of the form
\be
 \tilde\Phi_{n,\mathbf{k}}(r)\approx\frac{1}{r^{p+1}}A^+_{n,\mathbf{k}}+A^-_{n,\mathbf{k}},
\ee
where $A^{\pm}_{n,\mathbf{k}}$ are the constants of integration. The solution with the dependence $r^{-(p+1)}$ vanishes at the boundary and therefore is normalizable. Then, at large $r$ the zero mode $\Phi_0$ takes the form
\be 
\Phi_0(r,\mathbf{x})\approx\frac{L^{p+1}}{r^{p+1}}\phi_0^+(\mathbf{x})+\phi_0^-(\mathbf{x}).
\label{staticsolution}
\ee
Now that we have the form at the boundary of the static solution, let us consider an Euclidean action from which the equation can be derived. We propose 
\be
S=\frac{1}{2l^p}\int\ud^{p+2}X\sqrt{\det g}\, g^{MN}\pa_M\Phi\pa_N\Phi. 
\label{SEA}
\ee
For the normalizable zero mode, at large $r$, the on-shell action becomes
\be 
S_{\mathrm{on-shell}}^{(0)}\approx\frac{1}{2g^2}\int\ud^{p}\mathbf{x}\left[\la (\phi^+_0)^2+\pa_i \phi^+_0 \pa^i\phi^+_0\right], 
\ee
with
\be
\frac{1}{g^2}=\frac{1}{p+3}\frac{\b}{l^p}\frac{L^{p+4}}{r_i^{p+3}},\quad \la = (p+1)(p+3)\frac{r_i^2}{L^4}.
\label{couplingg}
\ee
The coupling $g$ and $\la$ depends on the characteristic scales of the system and the initial value of integration, as shown in Figure \ref{relscales}. We are only interested in the behaviour of the normalizable zero mode near to the boundary of the background. The radial integral is performed from $r_i$ to infinity and assumed that $r_i$ is close to the boundary. The reason behind this assumption is because the regime of large $r_i$ and close to the boundary implies that the coupling $g$ and $\la$ are strong. This will allow us to construct the gravitational dual of this field theory. After the field redefinition $\phi^+_0\to\phi^+_0/g$ we see that $\sqrt{\la}$ behaves as a mass term and the strong limit imply that the field $\phi^+_0$ is \emph{heavy}.

\subsection{Adding a $p$-brane and the $O(3)$ NL$\s$ model}
Let us add a $p$-brane located at $r=r_i$ into the Euclidean action given in Eq. \eqref{SEA}. Then
\begin{multline}
S= \frac{1}{2l^p}\int\ud^{p+2}X\sqrt{\det g}\, g^{MN}\pa_M\Phi\pa_N\Phi \\+ \frac{1}{2l^p}\int\ud^{p+1}x\sqrt{\c}\s, 
\label{SEA2}
\end{multline}
where $\c$ is the induced metric and $\s$  is the  tension of the $p$-brane. Taking the tension to be
\be
\s= -(p+1)\frac{L^{2p+1}}{r_i^{2p+2}}, 
\ee
the on-shell action for the normalizable zero mode will take the form
\begin{multline} 
S_{\mathrm{on-shell}}^{(0)}\approx\frac{1}{2g^2}\int\ud^{p}\mathbf{x}\left[\la\left( (\phi^+_0)^2-1\right)\right. \\ \left. +\pa_i \phi^+_0 \pa^i\phi^+_0\right].
\label{NSMac}
\end{multline}
Therefore we obtain a bounded potential in the strong limit if the the zero mode take its value on one of the two points of $S^0$.  The presence of the brane counter acts the heaviness of the field if it is constraint to be closed to the aforementioned points.  

This corresponds to the analogous situation of the strong coupling limit of the ``Mexican hat'' potential of a single scalar field $\psi$. The potential is of the form $V(\psi)=\la(\psi^2-f^2)^2$ where $\la>0$ and $f=\sqrt{m^2/\la}$ with $m^2>0$. Taking $\la\to \infty$ with the parameter $f$ fixed, the potential is bounded only near $S^0$ \footnote{ Moreover, by means of a Lagrange multiplier field $\mu(\mathbf{x})$, the potential in this limit can be written as a constraint $\mu(\mathbf{x})(\psi^2-1)$. We point out that in Eq. \eqref{NSMac} there is no need to introduce a Lagrange multiplier.}.

We can generalize Eq. \eqref{SEA2} in order to obtain a $O(3)$ NL$\s$ model. This can be achieved by exploiting the equivalence of the Lie algebra of $SU(2)$ and $O(3)$. We consider the scalar field to be in the adjoint representation of $SU(2)$ and the action is
\begin{multline}
S= \frac{1}{4l^p}\int\ud^{p+2}X\sqrt{\det g}\, g^{MN}\Tr\left[\pa_M\Phi\pa_N\Phi\right] \\+ \frac{1}{2l^p}\int\ud^{p+1}x\sqrt{\c}\s, 
\label{SEA3}
\end{multline}
 with $\Phi =\Phi^a T_a$, the set $\{T_a\}$ are the generators of $su(2)$. The on-shell action for the normalizable zero modes is
 \begin{multline} 
S_{\mathrm{on-shell}}^{(0)}\approx\frac{1}{2g^2}\int\ud^{p}\mathbf{x}\left[\la\left( \vec{\phi}^+_0\cdot \vec{\phi}^+_0-1\right)\right.\\ \left.+\pa_i \vec{\phi}^+_0\cdot \pa^i\vec{\phi}^+_0\right],
\label{NSMac2}
\end{multline}
where $\vec{\phi}^+_0\cdot\vec{\phi}^+_0=\phi^{+a}_0\phi^{+b}_0\d_{ab}$. In this approximation, the effect of adding a brane with negative tension and demanding that the potential should remain finite in the strong limit, is to compactify the target space (after adding a \emph{north pole}) of the normalizable  zero modes to $S^2$. 


\section{Gravitational dual }
\subsection{Spin-wave theory review}
First let us review the key results of spin-wave theory. The Hamiltonian of a ferromagnetic XXX Heisenberg  quantum spin chain is given by
\be
H=J\sum_{\langle i,j\rangle}\vec{S}_i\cdot\vec{S}_j, 
\label{HHam}
\ee
where $J<0$, $\langle i,j\rangle$ stands for the pair of nearest neighbour sites and $\vec{S}_i=(S_i^1,S_i^2,S_i^3)$. The spin operators $S^a$ (ignoring the lattice index) correspond to the generators of the Lie algebra associated to the Lie group $SU(2)$. They satisfy $[S^a,S^b]=i\epsilon^{abc}S^c$. A $(2s+1)$-dimensional (for $s=0,1/2,1,3/2,\ldots$) irreducible representation of $SU(2)$ has a basis $\{|s,m_s\rangle | m_s=-s,-s+1,\ldots,s-1,s\}$ such that
\bea
\vec{S}^2 |s,m_s\rangle &=& s(s+1)|s,m_s\rangle,\\
S^3|s,m_s\rangle &=& m_s|s,m_s\rangle,
\eea
and $S^{\pm}= S^1\pm i S^2$. The Holstein-Primakoff transformation is defined as
\bea
S^3&=&s-a^{\dagger}a,\\ 
S^{-} &=&\sqrt{2s}a^{\dagger}\sqrt{1-\frac{a^{\dagger} a}{2s}}, \\ 
S^{+}&=& \sqrt{2s}\sqrt{1-\frac{a^{\dagger} a}{2s}}\,a,
\eea
where the operators $a,a^{\dagger}$ satisfy $[a,a^{\dagger}]=1$. The vaccum state of the bosonic operators $|0\rangle$, is identified with the state $|s,m_s=+s\rangle$. In the $s\to\infty$ limit, the Hamiltonian given by Eq. \eqref{HHam}  becomes
\be
H=Js \sum_{\langle i,j\rangle}\left[a_ia^{\dagger}_j+a^{\dagger}_ia_j-a^{\dagger}_ia_i-a^{\dagger}_ja_j+s\right],
\label{H2F}
\ee
which corresponds to the semi-classical Hamiltonian in linear spin-wave theory. It can be diagonalized using the Fourier representation of the bosonic operators. Thus 
\begin{multline}
H=-\frac{1}{2}|J|s^2N\\+|J|s\sum_{\mathbf{k}}\left[\sum_{\mathbf{v}}(1-\cos(\mathbf{k}\cdot\mathbf{v})) \right]a^{\dagger}_{\mathbf{k}}a_{\mathbf{k}},
\label{HamilEFT}
\end{multline}
where $N$ is the number of sites in the lattice and $\mathbf{v}$ corresponds to the nearest-neighbourhood vector. In this Hamiltonian, the first term corresponds to the ground-state energy and the second term to the excitations referred to magnons. Quartic corrections of order  $O(1/s)$ to the Hamiltonian, Eq. \eqref{HamilEFT}, will introduce magnon-magnon interactions. The dispersion relation of a single magnon  is
\be
E(\mathbf{k})=|J|s \sum_{\mathbf{v}}(1-\cos(\mathbf{k}\cdot\mathbf{v})).
\ee
For small $\mathbf{k}\cdot\mathbf{v}$, we have $E(\mathbf{k})\sim |J|s||\mathbf{k}||^2$, which corresponds to a non-relativistic dispersion relation. In this low energy regime the magnons can be described by a $O(3)$ NL$\s$-model of the classical field $\vec{\phi}$ with $\vec{S}=s\vec{\phi}$ (due to the $s\to\infty$ limit) and $\vec{\phi}\cdot\vec{\phi}=1$. The $O(3)$ NL$\s$ model of magnons is assumed to be equal to the one given by Eq. \eqref{NSMac2}. By comparing the coupling $g$  defined in Eq. \eqref{couplingg} and the continuum limit of Eq. \eqref{HHam}, we identify
\be
|J|s^2 \leftrightarrow \frac{1}{6l^2}\frac{L^7}{r_i^6}. 
\label{Jr}
\ee
This result establishes the gravitational dual of the ferromagnetic magnons. 

\subsection{Ferromagnetic $J_1$-$J_2$  system}
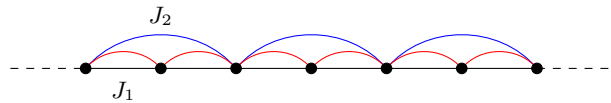
\begin{figure}[ht!]
\centering
\begin{tikzpicture}
\node  at (.5,-.3) {$J_1$};
\node  at (1,.7) {$J_2$};
\draw(0,0) -- (6,0);
\draw[dashed] (-1,0)--(0,0);
\draw[dashed] (6,0)--(7,0);

\draw[color=blue] (0,0) to [bend left=50] (2,0); 
\draw[color=blue] (2,0) to [bend left=50] (4,0); 
\draw[color=blue] (4,0) to [bend left=50] (6,0); 
\draw[color=red] (0,0) to [bend left=50] (1,0); 
\draw[color=red] (1,0) to [bend left=50] (2,0);
\draw[color=red] (2,0) to [bend left=50] (3,0); 
\draw[color=red] (3,0) to [bend left=50] (4,0);
\draw[color=red] (4,0) to [bend left=50] (5,0);
\draw[color=red] (5,0) to [bend left=50] (6,0); 
 \draw[fill=black] (0,0) circle (0.07);
 \draw[fill=black] (1,0) circle (0.07);
 \draw[fill=black] (2,0) circle (0.07);
 \draw[fill=black] (3,0) circle (0.07);
 \draw[fill=black] (4,0) circle (0.07);
 \draw[fill=black] (5,0) circle (0.07);
 \draw[fill=black] (6,0) circle (0.07);
\end{tikzpicture}
\caption{One dimensional representation of the system.}
\label{jjchain}
\end{figure}

The ferromagnetic $J_1$-$J_2$  spin chain with a competing interaction is defined by
\be
H=  J_1\sum_{\langle i,j\rangle}\vec{S}_i\cdot\vec{S}_j+J_2\sum_{\langle\langle i,j\rangle\rangle}\vec{S}_i\cdot\vec{S}_j, 
\ee
where $J_1,J_2<0 $ and $\langle\langle i,j\rangle\rangle$ denotes the next near-neighbour, as seen in Figure \ref{jjchain}. 

\begin{figure}[ht!]
\centering
\begin{tikzpicture}
\begin{axis}[
    xlabel = $\a$,
   xmin=-.5,   xmax=.5,
	ymin=0,   ymax=3,
	extra x ticks={0},
	extra y ticks={1},
	extra tick style={grid=major},
]
\addplot [thick,
   domain=-0.5:0.5, 
]
{1/(1+x)^6};
\addlegendentry{$\frac{|J_2|}{|J_1|}$}
\addplot [dashed, thick,
    domain=-0.5:0.5, 
]
{1/(1+x)^8};
\addlegendentry{$\frac{|\s_2|}{|\s_1|}$}
\end{axis}
\end{tikzpicture}
\caption{The $\alpha$-dependence of the ratio of couplings and the ratio of the brane tensions.  The ratios diverge for $\a\to-1$, as described in Equations \eqref{jratio} and \eqref{sratio}. }
\label{EpotPlot}
\end{figure}
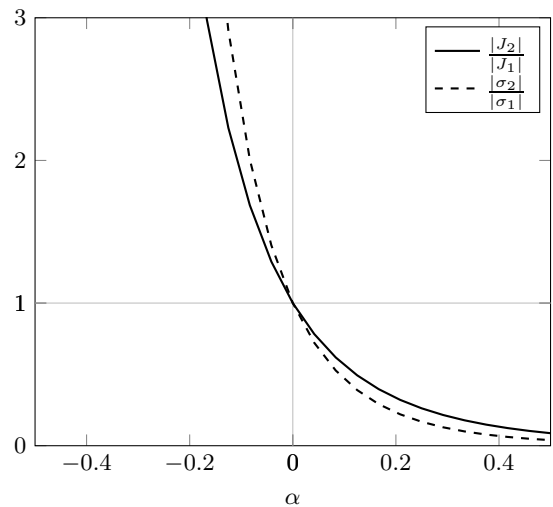

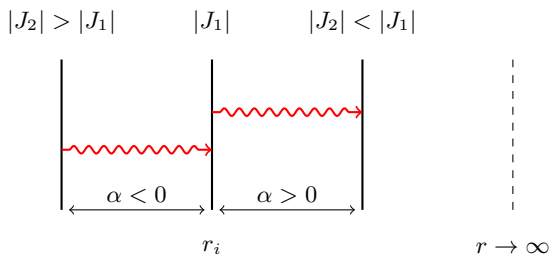
\begin{figure}[ht!]
\centering
\begin{tikzpicture}
\draw[thick] (0,0)--(0,2);
\draw[thick] (2,0)--(2,2);
\node at (2,-0.5) {$r_i$};
\draw[thick] (4,0)--(4,2);
\draw[dashed] (6,0)--(6,2);
\node at (6,-0.5) {$r\to \infty$};
\node at (0,2.5) {$ |J_2|>|J_1|$};
\node at (2,2.5) {$\small |J_1|$};
\node at (4,2.5) {$\small |J_2|<|J_1|$};
\draw[<->] (0.1,0)--(1.9,0);
\draw[<->] (2.1,0)--(3.9,0);
\node at (1,0.2) {$\a < 0$};
\node at (3,0.2) {$\a > 0$};
\draw[snakeline] (0,0.8)--(2,0.8);
\draw[snakeline] (2,1.3)--(4,1.3);
\end{tikzpicture}
\caption{Geometry of 3-branes and couplings. The fiducial heat flow between the branes is depicted. The absolute value of couplings behave as local fiducial temperatures.}
\label{branecoupgeo}
\end{figure}

The low energy dispersion relation of magnons gives 
$ E(\mathbf{k})\approx (|J_1|s+4|J_2|s) ||\mathbf{k}||^2$. Therefore, this suggest that we must add another scalar field $\Psi$ and a $3$-brane at a different location $\tilde{r}_i=r_i+\Delta r_i$. Let $S_2$ be the analogous action given by Eq. \eqref{SEA3} for the field $\Psi$ and $3$-brane. This action is of the form 
\be
S_2 =\int\limits_{r_i+\Delta r_i}^{\infty}\ud r\, L_2(r). 
\ee 
After the shift $r\to r-\Delta r_i$ and assuming $\Delta r_i$ to be finite and small, we can use a Taylor expansion for the Lagrangian. Therefore, up to $\Delta r_i$ corrections, the total action corresponds to the addition of the action for each field/$3$-brane system. Within this approximation the addition of systems is justified and by means of Eq. \eqref{Jr}, the couplings are related by
\be
\frac{|J_2|}{|J_1|}=\frac{1}{(1+\a)^6}, \quad \Delta r_i =\a r_i,
\label{jratio}
\ee 
where $\a$ parametrized the position of the second brane. The brane tensions are related by
\be
\frac{|\s_2|}{|\s_1|}=\frac{1}{(1+\a)^8}.
\label{sratio}
\ee
The $\a$-dependence of this ratios is depicted in Figure \ref{EpotPlot}.
On the other hand, the local fiducial temperature is given by
\be
T_{\mathrm{fid.}}(r)=\frac{T_H}{\sqrt{\frac{r^2}{L^2}h(r)}} .
\ee
For $r_i\gg r_+$, the gradient of fiducial temperature is
\be
\Delta T_{\mathrm{fid.}}\equiv \frac{T_{\mathrm{fid.}}(r_i+\Delta r_i)-T_{\mathrm{fid.}}(r_i)}{T_{\mathrm{fid.}}(r_i)}=\frac{1}{1+\a}-1. 
\ee
This expression suggests that the ratios given by Eq. \eqref{jratio}  with respect to the parameter $\a$ behave as the gradient of  local fiducial temperature. In particular, the absolute value of couplings behave as local fiducial temperatures 
\be
 \frac{|J_2|}{|J_1|}\leftrightarrow \left[ \frac{T_{\mathrm{fid.}}(r_i+\Delta r_i)}{T_{\mathrm{fid.}}(r_i)}\right]^6.
\ee
This is depicted in Figure \ref{branecoupgeo}.

\subsection{The action of the magnons as an effective action}
The action of the scalar/brane system, i.e. the action for the magnons, can be embedded in a  system in which gravity is dynamical. We are required to consider the partition function of a  gravity+matter action
\be
\cZ=\int\cD g_{MN}\cD\Phi\cD\Psi \exp\left(-S_{\mathrm{grav.}}[g]-S_m\right), 
\ee
where $S_m=S_m[g,\Phi,\Psi,\s_1,\s_2]$. The gravitational action $S_{\mathrm{grav.}}$ includes the Gibbons-Hawking-York surface term and the counterterm action (see \cite{Balasubramanian:1999re} and \cite{Elvang:2016tzz} for a recent perspective). The gravitons are introduced via the linear splitting of the metric $g_{MN}=\bg_{MN}+\kappa h_{MN}$, where $\bg_{MN}$ corresponds to the large SAdS-BH background and $\kappa^2 = 8\pi G_{p+2}$. Expanding the total action one finds
\be
\cZ=\ue^{-S_{\mathrm{grav.}}[\bg]}\int\cD\Phi\cD\Psi\, \ue^{-S_{\mathrm{eff}}} \int\cD h_{MN}\,\ue^{-\Delta S}, 
\ee
where $S_{\mathrm{grav.}}[\bg]$ is the finite gravitational on-shell action, $S_{\mathrm{eff}}=S_{\mathrm{eff}}[\bg,\Phi,\Psi,\s_1,\s_2]$ is the action used in the model and $\Delta S$ includes graviton interactions. In the weak gravity limit $\kappa\to 0$ the partition function becomes
\be
\cZ_{\mathrm{gravity+matter}}\approx \ue^{-S_{\mathrm{grav.}}[\bg]}\int\cD\Phi\cD\Psi\, \ue^{-S_{\mathrm{eff}}}.
\ee
If we expand $S_{\mathrm{eff}}$ around the normalizable zero modes, we obtain
\be
\cZ_{\mathrm{gravity+matter}}\approx \ue^{-S_{\mathrm{grav.}}[\bg]-S_{\mathrm{magnons}}}.
\ee
We find that the action of magnons corresponds to an effective theory. 

If we further assume that this partition function corresponds to the large limit of effective number of degrees of freedom ($c_{\mathrm{eff}}\to\infty$) of the partition function of string theory, the GKP-Witten relation in holography \cite{Witten:1998qj,Gubser:1998bc} tells us that $\cZ_{\mathrm{CFT}} =  \cZ_{\mathrm{gravity+matter}}$. In the AdS/CFT dictionary (see for example \cite{Aharony:1999ti,Klebanov:2000me,DHoker:2002nbb}), the non-normalizable solutions are regarded as the sources of the operators in the CFT and normalizable solutions correspond to physical fluctuations in the bulk. For the solution given in Eq. \eqref{staticsolution}, the field $\phi_0^-(\mathbf{x})$ corresponds to the source of an operator $\cO(\mathbf{x})$  of dimension $\Delta =4$ in the CFT. 

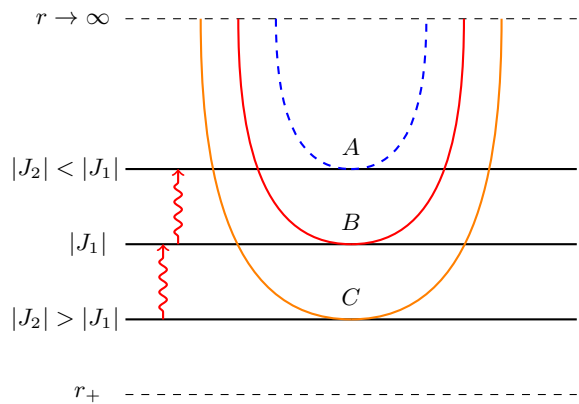
\begin{figure}[ht!]
\centering
\begin{tikzpicture}
\draw[dashed] (0,0)--(6,0);
\draw[thick] (0,1)--(6,1);
\draw[thick] (0,2)--(6,2);
\draw[thick] (0,3)--(6,3);
\draw[dashed] (0,5)--(6,5);

\draw[thick,blue,dashed] (2,5) to[out=-90,in=180]  (3,3);
\draw[thick,blue,dashed] (3,3) to[out=0,in=-90]  (4,5);

\draw[thick,red] (1.5,5) to[out=-90,in=180]  (3,2);
\draw[thick,red] (3,2) to[out=0,in=-90]  (4.5,5);

\draw[thick,orange] (1,5) to[out=-90,in=180]  (3,1);
\draw[thick,orange] (3,1) to[out=0,in=-90]  (5,5);

\draw[snakeline] (0.5,1)--(0.5,2);
\draw[snakeline] (0.7,2)--(0.7,3);
\node at (-0.5,0) {$r_+$};
\node at (-0.7,5) {$r\to \infty$};
\node at (-0.8,1) {$ |J_2|>|J_1|$};
\node at (-0.5,2) {$ |J_1|$};
\node at (-0.8,3) {$ |J_2|<|J_1|$};
\node at (3,1.3){$C$};
\node at (3,2.3){$B$};
\node at (3,3.3){$A$};
\end{tikzpicture}
\caption{Minimal surfaces in the planar SAdS-BH, the surfaces cannot penetrate the black hole horizon. The curves drawn in the figure are just illustrations. }
\label{entent}
\end{figure}

The planar SAdS-BH is dual to a CFT at finite Hawking temperature. We can relate the  spin chain couplings to holographic entanglement entropy at finite temperature in the CFT by relating the brane position with the turning point of minimal surfaces \cite{Rangamani:2016dms,Hubeny:2012ry}. Since the holographic entanglement entropy is proportional to the area of the minimal surface, from Figure \ref{entent}, we can deduce
\be
S_{C}>S_{B},\quad S_{B}>S_{A}. 
\ee
At high temperatures, finite temperature entanglement entropy reduces to the Stefan-Boltzmann law of entropy \cite{Calabrese:2009qy,Swingle:2011np}. The entropy  scales as $\Delta x \cdot T_H^3$ for $\Delta x \cdot T_H\gg 1$, where $\Delta x$ corresponds to the volume size of the subsystem in which the minimal surface is attached at the boundary. Hence, the ratios can be written as
\be
\frac{S_C}{S_B}=\frac{\Delta x_C}{\Delta x_B}>1, \quad  \frac{S_B}{S_A}=\frac{\Delta x_B}{\Delta x_A}>1,
\ee 
and we propose the relation between the couplings and entanglement entropy to be
\be
\frac{S_C}{S_B}\leftrightarrow \frac{|J_2|}{|J_1|},\quad   \frac{S_B}{S_A}\leftrightarrow \frac{|J_1|}{|J_2|},
\label{entropyratios}
\ee 
for $|J_2|>|J_1|$ and $|J_1|>|J_2|$ respectively.

\section{Summary and discussion}
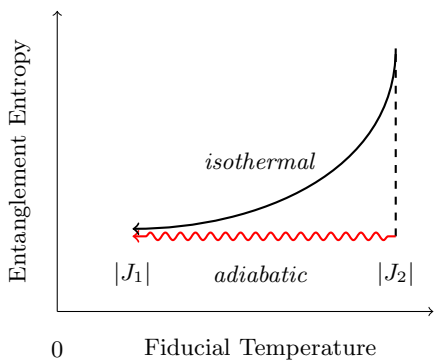
\begin{figure}[ht!]
\centering
\begin{tikzpicture}
\draw[->] (0,0)--(5,0);
\draw[->] (0,0)--(0,4);
\node at (0,-0.5) {$0$};
\node at (2.7,-0.5)  {Fiducial Temperature};
\node at (-0.5,2) [rotate= 90]{Entanglement Entropy };
\draw[thick, dashed] (4.5,1)--(4.5,3.5);
\draw[snakeline] (4.5,1)--(1,1);
\draw[->,thick] (4.5,3.5) to[out=-90,in=0]  (1,1.1);
\node at (1,0.5) {$|J_1|$};
\node at (4.5,0.5) {$|J_2|$};
\node at (2.7,0.5) {\emph{adiabatic}};
\node at (2.7,2) {\emph{isothermal}};
\end{tikzpicture}
\caption{\emph{Thermodynamic diagram} of the couplings for $|J_2|>|J_1|$.}
\label{thermo}
\end{figure}

The gravitational model of magnons of a ferromagnetic spin chain with a competing interaction in thermal equilibrium has been developed. Linear spin-wave theory allows us to interpret the couplings of the chain as an extra dimension in different positions. They correspond to the radial holographic coordinate in the planar limit of a \emph{large} SAdS-BH.
From the physical point of view, the consequences of these relations is summarized in the \emph{thermodynamic diagram} given in Figure \ref{thermo}. The \emph{isothermal} curve is at constant Hawking temperature in the CFT and the \emph{adiabatic} curve is realized in the bulk.
In order to study the \emph{thermodynamic diagram}, let us consider the classical version of the spin chain 
\bea
E&=&-|J_1|\sum_i\mathbf{S}_i\cdot\mathbf{S}_{i+1} -|J_2|\sum_i\mathbf{S}_i\cdot\mathbf{S}_{i+2}, \nn
&=&-s^2\left[|J_1|\cos\t+|J_2|\cos2\t\right]\sum_i 1,
\eea
The extrema of the energy follows from
\be 
\sin\t\left[|J_1|+|J_2|4\cos\t\right]=0.
\ee
The ferromagnetic ordered state solution is obtained with $\t=0$, a local maximum in the interval $\t\in[0,2\pi)$, without any relations between the couplings. The other possibility is
\be
\frac{|J_2|}{|J_1|}=-\frac{1}{4}\frac{1}{\cos\t}.
\ee
The ratio as a function of $\t$ reaches its minimum value  of $1/4$ at $\t=n\pi$ and blows up for $\t=\pi n/2,3\pi n/2$, where $n$ is a non-zero integer. The value $\t=\pi$ $(n=1)$ corresponds also to a local maximum of the energy.

This shows that the competing interaction allows the possibility of other state that is a local maximum of the energy  instead of a solely ferromagnetic ordered state. Nevertheless, as shown in subsection B of section III, the dispersion relation of the magnons of the system are quadratic in momenta even though the state is not ferromagnetic.

Due to the symmetric shape of the function, the angle domain is restricted to $\pi/2<\t\leq \pi$. The system is in a  incommensurable ordered state for  $\pi/2<\t< \pi$, i.e. the system is ordered  but without periodicity. For $\t_*\approx1.8234$ we get $|J_2|=|J_1|$. Therefore $|J_2|>|J_1|$ for $\pi/2<\t<\t_*$ and $|J_2|<|J_1|$ for $\t_*<\t<\pi$. 

For example, let us consider the case $|J_2|=2|J_1|$. This corresponds to $\t\approx 1.6961$. By means of  Eq. \eqref{jratio}, it is found that for this angle the second brane has two real values of the $\a$-parameter: $\a\approx -0.1901, -1.8909$. If instead we consider the reciprocal case $|J_2|=|J_1|/2$,  referred as the Majumdar-Ghosh point \cite{doi:10.1063/1.1664978}, we find only one real positive value $\a\approx 0.1225$. The point $\t=\pi$ has a real solution $\a\approx 0.26$ and the system is in a commensurate regime.

The range of physical significance for the $\a$-parameter is found to be $-1<\a\leq 0.26$. The closest brane to boundary saturates the upper bound and corresponds to the dashed minimal surface depicted in Figure \ref{entent}.  

In terms of the CFT at finite temperature in the regime $\Delta x \cdot T_H\gg 1$, the difference of entropy is  
\be
 S_C-S_B= \left(\frac{|J_2|}{|J_1|}-1\right)S_B, \quad  S_B-S_A= \left(\frac{|J_1|}{|J_2|}-1\right)S_A,
\ee
for $|J_2|>|J_1|$ and $|J_1|>|J_2|$ respectively, as seen in Figure \ref{entent}. In both cases, the difference of entropy is a positive semidefinite quantity and therefore it can be used to define a notion of \emph{distance} between  $|J_1|,|J_2|$ in a space of couplings. For the case $|J_1|>|J_2|$, the \emph{distance} reaches a maximum  of $3S_A$.

\subsection*{Acknowledgements }
We thank Marina David, Hrachya Khachatryan and Leopoldo P. Zayas for valuable discussions and comments for the improvement of the manuscript. 
\bibliographystyle{hsiam}
\bibliography{FMSAdSref} 

\end{document}